# Electronic transitions in Rydberg Matter as the source of the unidentified infrared bands (UIR) observed in interstellar space

Leif Holmlid, Atmospheric Science, Department of Chemistry, Göteborg University, SE-412 96 Göteborg, Sweden

Abstract.
Rydberg Matter (RM) theory is previously shown to give good agreement with the unidentified infrared bands (UIR, UIB) that are observed in emission from many objects in space. From stimulated emission studies in RM, a slightly simpler theoretical model is found that describes the transitions giving the UIR bands better. The general theory for RM was recently shown to be very accurate using rotational spectroscopy to observe several types of RM clusters (Mol. Phys. 105 (2007) 933). Two different large studies of UIR are now interpreted with the improved model. High-resolution UIR observations by Beintema et al. are interpreted as transitions from high levels $n$ = 40 - 80 down to levels $n$ = 5 - 12 in the final state. The upper level quantum numbers for the observed bands agree well with those found in experiments. The stationary UIR bands at 3.3 and 11.3 μm are due to deexcitation $n = 9 \rightarrow 5$ and $9 \rightarrow 7$ of Rydberg species through stimulated emission. The other UIR bands vary considerably in wavelength in different environments and even with time as seen in numerous observational studies. Their great variability is difficult to model by the commonly used PAH model, but is a natural consequence of the varying excitation state of RM.

# 1. Introduction

Recent results on stimulated emission spectroscopy of Rydberg Matter (RM) (Badiei & Holmlid 2003, 2005; Holmlid 2004b) now make it possible to complete a more accurate comparison with the so called unidentified infrared bands (UIR bands or UIB), since the transitions involved are better understood than in our previous study (Holmlid 2000). That first report on RM as the source of the UIR bands was mainly based on theory, but the experimental studies of RM can now also be directly compared with observations. The results of immediate interest are the stimulated emission spectroscopy studies of RM published recently (Holmlid 2004b; Badiei & Holmlid 2005). The excitation level in RM can be determined from the spectra and the transitions can be identified from the bands measured in stimulated emission. The general theory for RM was recently shown to be very accurate using rotational spectroscopy to observe several types of RM clusters in the radio frequency range (Holmlid 2007b).

A large number of studies have observed the UIR bands from sources as diverse as supergiants (Sylvester et al. 1994), novae (Evans et al. 1997) and the whole Galaxy (Kahanpää et al. 2003) besides the many other objects reported and summarized prior to 1997 (Geballe 1997; Tokunaga 1997). The "canonical model" for the origin of the UIR bands at present is that large densities of maybe a hundred different types of PAH molecules exist in interstellar space. However, the large density of carbon required for this means that a "carbon crisis" is observed to exist (Dwek 1997), with too little carbon available in space. Several other arguments against PAH molecules as the source of UIR emission exist, as described below. Relatively few spectroscopic studies of emission from PAH molecules have been made, and they do not agree well with the UIR bands. The most active research group in this field concludes that neutral PAHs are not likely to be a major contributor to the UIRs (Kim & Saykally 2002). That RM is an important form of matter in the ISM can be concluded from a few previous studies. The UIR bands were proposed to be formed by RM by Holmlid (2000) with good agreement between predicted and observed band positions. Detection of similar bands in RM by Raman spectroscopy confirmed that this proposal was valid (Holmlid 2001b). Maybe the most detailed and striking confirmation of RM in interstellar space is the accurate calculation of more than 60 of the so called diffuse interstellar bands (DIBs) as due to transitions in RM (Holmlid 2004a). The approximately 280 DIB bands have evaded interpretation for more than 80 years since their first observation, but a model is now developed that explains practically all the bands. It may also be interesting that spectroscopic results from comets can be interpreted well as due to RM (Holmlid 2006a); in fact, several bands observed in comets are identical to those identified as UIR bands in interstellar space. The alkali atmospheres on the Moon and Mercury are also recently proposed to have the form of RM clouds (Holmlid 2006b).

RM is further proposed to explain several features in intergalactic space. The intriguing observations of Faraday rotation effects at radio frequencies in intergalactic space (Clarke et al. 2001) were interpreted quite satisfactorily with the RM model (Badiei & Holmlid 2002c). This means that a low density of RM exists in intergalactic space. The quantized redshifts observed for a large number of galaxies (Guthrie & Napier 1996) are interpreted as stimulated Raman shifts in intergalactic RM (Holmlid 2004c). These results show that RM can be considered seriously as a source of spectroscopic features even in intergalactic space.

Some features of RM of interest for the astrophysical environment will be described first, followed by the main results from the comparison of two sets of UIR observations with theory. In the discussion section, the great variability of the UIR bands is shown to be well described by the RM model.

## 2. Background

The physics of RM is described in several recent publications on for example the stimulated emission in the infrared (Badiei & Holmlid 2003, 2005; Holmlid 2004b) and in a few astrophysical applications (Badiei & Holmlid 2002b, 2002c; Holmlid 2004c). The description of RM in Holmlid (2004a), concerned with the diffuse interstellar bands (DIBs), may be the most relevant for the present report. The behaviour of RM in space will here only be described briefly. H atoms is probably the main constituent of the RM clusters in space, and this type of matter has recently been shown to exist (Badiei & Holmlid 2004a, 2004b), even in a form that has the same bonding distance as metallic hydrogen. Also $H_2$ molecules and most other atoms and small molecules will take part in the formation of the RM clusters. The bonding distances between the atoms and molecules in undisturbed RM clusters at an excitation level of $n = 80$ are 1 μm, and the diameters of the clusters are a few μm. Thus, they are of the same size as deduced for many grains in space. All the atoms (molecules) in such a cluster normally have the same excitation state $n$, as required for the coherent motion of the RM electrons and the bonding in the RM (Holmlid 1998).

The formation of RM clusters in space involves desorption from other particles, especially graphite and metal oxide particles. For this, only a low temperature is needed, and RM of $H_2$ molecules can be formed by desorption from graphite and metal oxide surfaces in the laboratory in a vacuum already at 300 K (Wang & Holmlid 2002; Wang et al. 1999). Several types of experimental studies show the formation of RM and RM clusters at surfaces of carbon and metal oxide. These results have been independently confirmed (Kotarba et al. 2000a, 2000b, 2001). It is expected that the formation of RM should be facile for example outside carbon emitting stars with particle formation in the outgoing flow at several hundred K.

The question of stability in space of RM is of great interest. Manykin et al. (1981, 1992a, 1992b) calculated the lifetime of RM in different excitation levels, and found it to be relatively long, of the order of 100 years at an excitation level of $n = 16$. At $n = 80$ in interstellar space, the lifetime from a simple extrapolation would be extremely long according to this calculation, longer than the lifetime of the universe. Thus, RM can be formed at an extremely low rate and still occupy large volumes of interstellar space.

When RM has formed, it will be excited and thus maintained by energetic and ionizing radiation. This is so since Rydberg states are formed by recombination of electrons and ions, and RM is in principle formed by the condensation of circular Rydberg states. Energetic processes will deposit energy in the RM by recombination giving high Rydberg states. These states add on to existing RM clusters. The contact of RM clusters with ground state atoms and molecules gives incorporation of the ground state species into the clusters after sharing of excitation energy. This is observed in our laser fragmentation time-of-flight experiments (Badiei & Holmlid 2004a, 2004b; Wang & Holmlid 2002) as the formation of clusters with masses corresponding to mixed clusters.

In the laboratory, the RM clusters formed survive intense focused laser light in the form of ns pulses with up to $10^{10}$ W cm$^{-2}$ intensity (Wang & Holmlid 1998, 2000). Most parts of space where RM is observed will probably be much calmer than the laboratory vacuum chamber with such laser intensity, so the stability of RM is sufficient for it to survive for a long time in interstellar space. RM is neither destroyed by intense IR radiation. On the contrary, strong IR fields promote RM cluster re-excitation as shown directly by the successful operation of the RM laser (Badiei & Holmlid 2003, 2005; Holmlid 2004b). The temperature range 300 -1300 K is used to form RM in the laboratory. This range covers the temperatures in space where the UIR emission originates.

The emission process of interest in the present study is shown in Fig. 1 in the form of a de-excitation process in RM formed from atoms or molecules M. Single-electron emission processes are in principle forbidden since angular momentum can not be conserved in transitions from circular high to low Rydberg states (or between the similar states in the condensed phase), which means that further electrons in the condensed material take part in the transition. Thus, we have previously proposed a model where another core electron is involved (Holmlid 2000, 2004b), so that the total emission process is a two-electron process. This process may in the case of a transition starting at a high level ($n_2\, l_2$) in the conduction band be described as

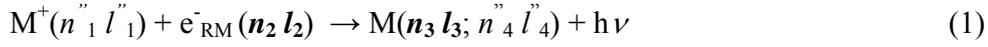

$$M^+(n''_1\, l''_1) + e^-_{RM}(n_2\, l_2) \rightarrow M(n_3\, l_3;\, n''_4\, l''_4) + h\nu \qquad (1)$$

where the double primes indicate quantum numbers with effective core ion charge $Z = 2$, thus for orbits in an ion. The bold type indicates the main energy change approximately corresponding to the emitted photon. The electron in the core ion is characterized by quantum numbers $n''_1\, l''_1$ initially, and $n''_4\, l''_4$ finally, and this change is in the model proposed to cover the angular momentum change of the de-excited electron going from $n_2\, l_2$ to $n_3\, l_3$. However, this model requires that the electron giving off the main energy for the photon starts in the conduction band. In Holmlid (2004b) the upper, most likely excitation level was derived from the band peak positions, for $n'' \geq 9$. The upper level was found to be $n'' = 40 - 80$, thus at levels that obviously must exist in the conduction band. However, for transitions to $n'' < 9$ the electron often originates in levels much too low to correspond to the conduction band. This is shown for example in Table 2 in Holmlid (2004b). Thus, these transitions seem to take place more or less completely within the core ion itself. A reasonable description could be $M^+(n''_1\, l''_1) \rightarrow M^+(n''_4\, l''_4) + h\nu$, but the angular momentum conservation is then not explicitly included. A more complete description is

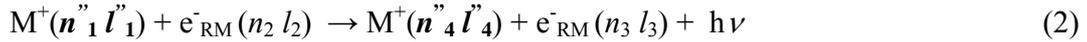

$$M^+(n''_1\, l''_1) + e^-_{RM}(n_2\, l_2) \rightarrow M^+(n''_4\, l''_4) + e^-_{RM}(n_3\, l_3) + h\nu \qquad (2)$$

where the conduction band electrons take care of the angular momentum conservation. In Eq. (2), the final state is still the intact RM, while in Eq. (1) the final state was an atom within the RM phase. In Eq. (1), transitions down to $n''$ odd are somewhat difficult to imply: in this case which is often observed in experiments, Eq. (2) is more useful.

Further points concerning this type of transition are discussed in Holmlid (2007a). Here, it is sufficient to describe the energies and wavelengths for the band centers by

$$\tilde{\nu} = 1/\lambda = R(1/n_3^2 - 1/n_2^2) \qquad (3)$$

where $R$ is the Rydberg constant. This formula is a reasonable approximation based on experiments for the spectral peaks. It is apparent from Eq. (1) that the emission will be in the form of broad bands, since the electrons $e^-_{RM}$ are in the conduction band in RM and the width of the emission band will correspond at least to the height of this conduction band. In the case of Eq. (2), all energy changes of the coupled electrons will be reflected in the size of the finally emitted quantum. Thus, also in this case the transitions are expected to give relatively broad bands, but probably not as broad as for the process in Eq. (1). The corresponding transitions will take place in any atom or small molecule but H atoms.

## 3. Results

To make an accurate comparison between RM theory and observations, it is necessary to chose a set of UIR bands from a study where many different peaks are measured with a high precision. Beintema et al. (1996) used the SWS of the Infrared Space Observatory (ISO) on three different C-rich stars and measured a large number of bands in each spectrum with high resolution. In Table 1, all peak values are shown that were found by them. The most intense bands as given by Beintema et al. are in bold print. The corresponding transitions for all peaks have been calculated from Eq. (3) and are included in Table 1 as the upper and lower quantum numbers for the transitions. The values $n_2$ are plotted against wavelength in Fig. 2 with the strongest bands marked by circles. The first point to observe from this figure is that the series for $n_3 = 6$ and 11 display peaks with high values of $n_2$, close to the ionization limit. This behaviour with $n_2 > 50$ is unexpected for disturbed RM, which in general has lower $n_2$ values as seen from the other results. This effect is probably due to the interaction with ordinary Rydberg atom transitions at 3.296 and 11.306 μm. See further the discussion below. The second point to note is that the most intense bands appear either at the lowest $n_2$ value (normally from NGC 7027) or at intermediate values from the two other stars, in a regular fashion. This would indicate a higher level of excitation in the two other stars relative to NGC 7027. The observed trend that $n_2$ increases with $n_3$ is expected, as an effect of overlap of the upper and lower states. Thus, the observed peaks fall nicely into the RM pattern. It should be observed that not all wavelengths in the range studied could have been interpreted as due to transitions in RM: this is the reason why the values 3.296 and 11.306 μm have to be singled out as due to a slightly different process. Further, there are gaps correctly between the different $n_3$ assignments, as seen in Fig. 2.

The UIR peaks measured by Kahanpää et al. (2003) for the interstellar medium across the Galaxy are given in Table 2. These data are interesting since they should correspond to an average for all UIR emitting locations in space. The authors have also observed a major peak at 14.1 μm, that is not observed by Beintema et al. Most observations fall on a smooth curve in Fig. 3 with reasonable values of $n_2$, in the same range as the experimental data, i.e. $n_2 = 20 - 50$. The experimental results from Holmlid (2004) are also included in Fig. 3. The results for $n_3 = 6$ and 11 are divergent due to interaction with emission from Rydberg species, as described above and in the discussion. With the new RM model for the transitions, three of the main peaks measured by Kahanpää et al. are interpreted as due to transitions from values of $n_2$ lower than 30, as seen in Table 2. The agreement between observations, experiments and theory is satisfactory.

## 4. Discussion

Since there at present exists a "canonical" interpretation of the UIR bands in the form of rovibrational transitions in PAH molecules, the PAH model is also discussed below. The RM model used here is, contrary to the PAH model, supported both by theory and experiments; however, a further discussion of the variability of the UIR bands is also given in the final subsection. Here, the main point is that such a shifting character of the spectra as observed is highly unlikely to be found even with a large number of different PAH molecules involved; instead, this behaviour is naturally expected in the RM model.

## 4.1. Interference with Rydberg transitions

RM is built up of Rydberg species, and under intense particle and photon impact, the RM will be partly fragmented to smaller particles, even Rydberg atoms, as seen in our laser fragmentation experiments (Wang & Holmlid 2002, 2002; Badiei & Holmlid 2002a, 2002d). Thus, relatively low Rydberg states of atoms and small molecules like $H_2$ will exist in contact with the RM. Their radiative lifetime is relatively long, and they may deexcite by stimulated emission processes. If such stimulated transitions coincide with UIR transitions in the RM, relatively sharp emission peaks may result. Relevant wavelength values for Rydberg transitions are given in Table 3. A comparison reveals that accidental coincidences exist at a few positions, especially at 3.296 and 11.306 μm, at transitions $n = 9 \rightarrow 5$ and 7 in Rydberg species respectively (bold type in the table). Thus, the strong UIR peaks at 3.287 - 3.296 and 11.200 - 11.260 μm in Table 1 are likely due to enhancement of the UIR signal by stimulated emission from Rydberg species. This explains the departure from the general trends at $n = 6$ and 11 in Figs. 2 and 3. It may be observed that the same situation exists for several IR bands observed from comets, especially for the bands close to 3.296 and 11.306 μm. For a more complete description see Holmlid (2006a).

## 4.2. Comparison with experiment

The agreement between experimental and observational transitions in Fig. 3 is encouraging, considering that the processes of band formation are slightly different in the case of mainly spontaneous emission in space giving the observed UIR, and stimulated emission in a cavity in the experiments. This was described briefly in section 2. Other differences of course exist. For example, the distributions of core ion masses are different in the two cases, being mainly H and $H_2$ in space and K and $N_2$ in the experiments. The conditions for formation of RM are also different, at 1000 K emitter temperature and 300 K of the surrounding chamber in the experiments, with no energetic particle fluxes. A hot surface is only a few mm away, giving rapid re-excitation of the clusters after stimulated emission has taken place. The stability of large clusters will also be influenced by the photon flux. In space, the temperature is often cited to be 400 - 700 K (Evans et al. 1997), and fluxes of energetic radiation like UV photons exist. The direct IR flux may be lower than in the laboratory. It might be expected that a correlation should exist between the $n$ values, such that a high excitation value $n_2$ for the initial state would give a relatively high value of excitation $n_3$ for the final, lower state. The reason for this is the overlap, which would simplify such transitions. Some of this dependence is seen in Fig. 3, with positive slope of the curve found from the observations when points at $n_3 = 6$ and 11 (both due to Rydberg species emission) are excluded. In the laboratory experiments, a slightly different trend is observed. However, the overall agreement in trends and especially in the excitation level $n_2$ is satisfying.

## 4.3. Variation of UIR bands in observations

To complete the discussion and the application of RM to the UIR problem, a comparison with studies of UIR bands that are variable in wavelength with time or position in space is presented here. Such quite common observational results indicate that the UIR bands are not fixed in wavelength as they should be if they are due to vibrational transitions, like in PAH molecules. If they instead depend on the excitation level in RM that varies with the environment where the UIR bands are observed, such variations are easily understood. Already the fact that the UIR spectra are variable for different objects points in this direction. The intensity ratios of the various UIR bands are fairly constant in many studies, like in the studies of UIR bands in the interstellar medium by Chan et al. (2001). That the intensity of the UIR bands vary systematically (increasing linearly) with the integrated radiation intensity in space (Onaka 2000) is in good agreement with the RM model, where the excitation process is directly coupled to the emission process.

One example of variable UIR features is provided by the study of Nova V705 Cas (1993) by Evans et al. (1997). The authors observe the variation in time of the UIR bands from the nova for up to 750 days after its maximum light. In the N-band around 10 μm, they observed shifts of the peak at 8.2 μm to 8.0 μm during less than 90 d. They conclude that this band is the one normally observed at 7.7 μm, in the RM model assigned to $n_3 = 9$. That an evolution in time is found means a small change of the upper excitation level $n_2$ from 29 to 32, which could indicate a stabilization of the RM phase at lower temperatures in the outflow from the nova. After 600 d, the peak has shifted to 7.7 μm, i.e. to a higher level $n_2$ in the RM of 45. This type of behaviour is expected for RM where lower densities and less disturbing environments will give higher excitation values in the RM; this effect can hardly be explained in the PAH model.

Not all UIR bands were investigated by Evans et al. (1997). However, the time evolution of the 3.3 and 3.4 μm bands was also followed. The 3.3 μm band was strongly peaked 251 d after maximum light, but became broader and less peaked with time, but with no shift in wavelength. As discussed above, the sharp peak at 3.3 μm is mainly due to a Rydberg transition $n = 9 \rightarrow 5$ by stimulated emission, and this peak should thus not change in position. In the RM model it is further enhanced by the optical gain in the RM. The broader band at 3.4 μm corresponds to an RM transition with an upper level at $n_2 = 33$, similar to values deduced for the 8.0 μm band above. This band shifts slightly in wavelength with time. Thus, two different types of processes exist for these two bands.

To conclude, the formation of Rydberg species and RM in the material ejected from the nova studied by Evans et al. is highly likely. On the other hand, that large free PAH molecules should be formed rapidly and exist in large amounts in the strong UV flux is unlikely.

An extreme case where UIR type bands are observed but PAH molecules should not be able to exist is the hydrogen-deficient Wolf-Rayet star studied by Chiar et al. (2002). They observe bands peaking at 6.4 and 7.9 μm. Based on the RM description, the corresponding transitions are from $n_2 = 27$ and 35 down to $n_3 = 8$ and 9 respectively. The small $n_2$ values show that the average excitation level of RM is very low in this case, probably due to the harsh environment with strong particle fluxes and high effective temperatures. The observations are thus understood in the RM model, but are at variance with the PAH model.

A very detailed and thorough study of the UIR bands was recently made by Peeters et al. (2002). Their results show that many correlations exist between the different band positions and subdivisions into other bands, believed to be located at specific wavelengths. Overall, their study shows the large variability of the bands both in position and in shape. This is true especially for the 6.2, 7.7 and 8.6 μm bands. In the case of the 3.3 and 11.3 μm bands they were found not to vary in wavelength. They correlate in intensity, but do not correlate well in intensity with the other bands. As described above, this is due to the main origin of these two bands in Rydberg species, while the other UIR bands are due to transitions in RM.

The classification of the UIR emitting objects in class A, B and C done by Peeters et al. (2002) shows that the 6.2, 7.7 and 8.6 μm bands all increase in wavelength in the order A, B and C. In the RM description, the differences in wavelengths are mainly due to the excitation level in the RM, being highest in class A objects. The common feature of class A objects is that they are more highly dispersed and more stable regions (HII regions, galaxies etc.) and thus have larger excitation levels (large interionic distances) in the RM. This is as expected for RM. Classes B and C on the other hand are denser objects and even with quite newly formed dust atmospheres, thus at lower excitation levels since the binding energy in the RM is higher for low excitation levels; they are not so easily fragmented by particle and UV impact. The classification thus agrees with the RM model.

A study by Lambert el al. (2000) seems to find that UIR features are only observed in one of the three R Coronae Borealis stars studied. The UIR features were only observed in the star which has the highest hydrogen density. The two others are hydrogen-poor. However, that study concentrates on the peaks in the IR distributions and not on the total emission. All three stars show broad IR bands that seem to overlap due to relatively high temperatures of the emitting medium. The UIR bands reported, at 3.294, 6.298, 11.280 and 13.490 μm, are relatively sharp, even if the fourth one does not seem very sharp in the spectrum shown by them. The first three bands are very close to Rydberg transitions in Table 3, namely 3.296, 6.290 and 11.306 μm. Thus, one can conclude that the sharp bands are due to Rydberg species while the broad UIR bands are due to general RM de-excitation.

A final example of the variability of the UIR features is provided by the Red Rectangle in several studies. Mékarnia et al. (1998) observed the nebula at wavelengths coinciding with several predicted but usually not studied RM bands, at 2.20 ($n_3$ = 5), 3.30 (6), 3.87 (possibly $n_3$ = 6) and 4.75 (7) μm. The intensity contour plots for the 3.30 μm filter was different from (more spiked than) the other plots which were all quite similar. Kerr et al. (1999) observed that the 3.3 μm band did not correlate with the other UIR emission bands. A similar result was found by An & Sellgren (2003). They compared the 3.29 μm band with the corresponding continuum around 2 μm, and conclude that these two features have different sources, emission or formation mechanisms. As discussed above, it is concluded here that the 3.3 and 11.3 μm bands have strong contributions from Rydberg species deexcitation through stimulated emission. Thus, there will just be an "indirect link" (Kerr et al. 1999) between the 3.3 μm feature and other UIR bands from RM.

It is concluded that overwhelming evidence shows that the UIR bands vary strongly with the environment, even rapidly in time, which is easily described as changes in excitation level in RM but very difficult to predict from the PAH model. In the same way, much evidence exists showing that the common 3.3 and 11.3 μm bands behave differently from the other UIR bands. This is due to their origin in Rydberg transitions. (The same emission with bands at 3.3 and 11.3 (11.2) μm is also known from IR studies of comets (Holmlid 2006a). It is even noted from observations (Hanner et al. 1997) that the 11.2 μm band does not shift with varying conditions, while other IR bands from comets like the one at 10.0 μm do shift considerably.) That the UIR bands appear quite similar in many objects studied is thus due to their common origin in RM, not to an unknown mix of charge states and molecular compositions of PAH molecules.

## 4.4. The PAH model

The UIR bands have for some time been considered to be related to PAH molecules and other carbonaceous compounds in interstellar space. The large variability of the UIR features discussed above gives valid proof against this relation, but the large number of possible molecules and possible charge states has made the PAH suggestion quite long-lived. However, it is of course not enough to compare absorption spectra of PAH molecules with UIR bands, since molecular spectra are different in emission and absorption. The problem with the influence of the actual internal temperature of the PAH molecules on the emitted spectra, for example how broad the bands are, has generally not been taken into account in PAH studies. Calculations of spectra of such complex molecules do not give very accurate predictions of the experimentally observed bands, as seen for example in the results from Pauzat & Ellinger (2002). The experimental studies by Saykally et al. are thus extremely valuable contributions in this field, giving emission spectra from neutral PAHs (Wagner et al. 2000) as well as from some PAH cations (Kim et al. 2001). Their conclusion is that neutral PAHs are not likely to to be a major contributor to the UIRs (Kim & Saykally 2002), but they find that some ion forms agree better with actual UIR spectra. The search for more complex molecules to give reasonable spectra is still continuing. When such spectra are found, it is still necessary to 1) describe the formation of the spectrum through excitation by UV light or particle impact on these PAH molecules, 2) verify that a reasonable thermal excitation gives the broad UIR band structure by their rotational band structure, 3) verify that the PAH molecules of interest do not dissociate in the UV and particle flux intensity, 4) describe the processes leading to the formation of just these types of molecules and not the full range of PAHs, 5) describe the processes of forming the PAH molecules in space from the atoms and smaller molecules present in all the various environments, 6) clarify why these molecules seem to have so variable spectra in various environments, 7) describe why no other molecules in interstellar space emit more intensely than these PAH molecules with their multitude of transitions; the intensity of each band will be very low.

In comparison with the PAH model, RM as the UIR precursor has many advantages: 1) it is in a long-lived excited state, not directly dependent on the thermal environment, 2) the excitation channel by IR, UV and particle impact is direct, giving the emitting state, 3) all matter in interstellar space (but large molecules) can take part in the RM formation, and there is no strong requirement on the types of atoms involved, 4) RM is likely to exist in most parts of space, probably being the must abundant state of matter in the universe, 5) the spectra will change with excitation state, thus with environment in a rather easily predictable fashion (see above), 6) the UIR emission from RM is the main process of deexcitation.

Besides, the RM model by its pure simplicity is preferable to the PAH model, which has been studied for more than 20 years without a positive agreement. Negative test outcomes exist, on the other hand. For example, the detection of UIR features in the spectrum of a hydrogen deficient Wolf-Rayet star (Chiar et al. 2002) disproves that these features are due to PAHs. A study of UIR bands over the Galaxy (Chan et al. 2001) shows that there is no systematic variation of the ratios of several UIR bands over large parts of the Galaxy, with no dependence on the UV radiation field strength. It also shows that any PAHs giving the UIR bands must be mainly ionized even at low UV intensities. This is not in agreement with theoretical studies that show that a large fraction of the PAHs is predicted to be neutrals. It is then also an open question why the neutral PAHs do not emit in the IR and are observed in the UIR spectra, as their cations are assumed to do.

## 5. Conclusions

The improved RM theory predicts bands in agreement with high resolution observations of the UIR bands from circumstellar matter and with observations of UIR bands from the Galaxy. The transitions studied experimentally by stimulated emission give quantum numbers similar to the observed bands. The variability of the UIR band wavelengths is well described by RM theory. One example is in the outflow from Nova V705 Cas where the UIR bands change in position, drifting to shorter wavelengths with time on the order of months. This is easily described by the RM model, but cannot be understood in the PAH model. The UIR bands at 3.3 and 11.3 μm do not change in wavelength as the other bands do. They coincide with transitions $n = 9 \rightarrow 5$ and $9 \rightarrow 7$ in Rydberg species, and are observed through stimulated emission processes amplified by RM transitions.

Table I. Assignments of the UIR bands observed by Beintema et al. (1996). The main columns show the observed peaks in three different stars. The upper and lower principal quantum numbers are given for the transitions $n_2 \rightarrow n_3$ in the IR range from 3 to 19 μm. Numbers in bold type are the most intense bands.

| HR 4049 | | | IRAS21282+5050 | | | NGC 7027 | | |
|---|---|---|---|---|---|---|---|---|
| λ (μm) | $n_2$ | $n_3$ | λ (μm) | $n_2$ | $n_3$ | λ (μm) | $n_2$ | $n_3$ |
| 3.248 | 9 | 5 | 3.248 | 9 | 5 | 3.287 | 9 | 5 |
| 3.296 | 90 | 6 | 3.294 | 97 | 6 | | | |
| 3.342 | 45 | 6 | | | | | | |
| | | | 3.397 | 33 | 6 | 3.398 | 33 | 6 |
| 3.423 | 30 | 6 | 3.435 | 29 | 6 | 3.467 | 26 | 6 |
| | | | 3.478 | 25 | 6 | 3.481 | 25 | 6 |
| 3.531 | 23 | 6 | 3.538 | 23 | 6 | 3.518 | 23 | 6 |
| | | | | | | 3.546 | 22 | 6 |
| | | | 3.569 | 21 | 6 | 3.574 | 21 | 6 |
| 3.757 | 17 | 6 | 3.633 | 20 | 6 | 3.624 | 20 | 6 |
| | | | | | | 5.234 | 19 | 7 |
| | | | | | | **5.650** | 16 | 7 |
| 6.010 | 47 | 8 | 6.006 | 47 | 8 | | | |
| | | | **6.217** | 32 | 8 | **6.219** | 32 | 8 |
| **6.258** | 31 | 8 | | | | | | |
| 6.318 | 29 | 8 | | | | | | |
| | | | 6.940 | 20 | 8 | **6.924** | 20 | 8 |
| 7.602 | 53 | 9 | 7.609 | 52 | 9 | 7.594 | 54 | 9 |
| | | | **7.724** | 43 | 9 | | | |
| **7.892** | 36 | 9 | | | | **7.815** | 39 | 9 |
| 8.068 | 31 | 9 | | | | | | |
| 8.593 | 24 | 9 | **8.626** | 24 | 9 | **8.585** | 24 | 9 |
| **8.679** | 23 | 9 | | | | | | |
| | | | 9.533 | 48 | 10 | | | |
| 9.707 | 41 | 10 | | | | | | |
| | | | | | | **10.44** | 28 | 10 |
| 11.04 | 24 | 10 | 11.05 | 24 | 10 | 11.05 | 24 | 10 |
| | | | **11.23** | 82 | 11 | **11.22** | 85 | 11 |
| **11.26** | 77 | 11 | 11.30 | 74 | 11 | | | |
| 11.67 | 47 | 11 | | | | 11.76 | 44 | 11 |
| 11.88 | 41 | 11 | **12.05** | 38 | 11 | 11.95 | 40 | 11 |
| 12.75 | 30 | 11 | **12.63** | 31 | 11 | **12.73** | 30 | 11 |
| | | | 13.48 | 26 | 11 | 13.58 | 25 | 11 |
| | | | **16.99** | 25 | 12 | **17.03** | 25 | 12 |
| | | | 18.02 | 23 | 12 | 17.86 | 24 | 12 |

Table II. The transitions $n_2 \rightarrow n_3$ observed by Kahanpää et al. (2003) in the IR range from 3 to 15 μm.

| $n_3$ | λ (μm) | $n_2$ | λ (μm) | $n_2$ |
|---|---|---|---|---|
| 6 | 3.3 | 78 | | |
| 7 | 5.25 | 18 | | |
| 8 | 6.2 | 33 | 6.9 | 20 |
| 9 | 7.7 | 44 | 8.6 | 24 |
| 10 | 9.7 | 41 | | |
| 11 | 11.3 | 71 | 12.7 | 30 |
| 12 | 14.1 | 46 | | |

Table III. Transitions in separate Rydberg species using the Rydberg constant for hydrogen molecules, in the range 2 -20 μm. Transitions observed in the UIR emission are in bold type.

| $n_{low}$   $n_{high}$ | 14 | 13 | 12 | 11 | 10 | 9 | 8 | 7 | 6 | 5 |
|---|---|---|---|---|---|---|---|---|---|---|
| 10 | 18.610 | | | | | | | | | |
| 9 | 12.584 | 14.179 | 16.876 | | | | | | | |
| 8 | 8.662 | 9.389 | 10.501 | 12.384 | 16.205 | | | | | |
| 7 | 5.955 | 6.290 | 6.770 | 7.506 | 8.758 | **11.306** | 19.057 | | | |
| 6 | 4.020 | 4.170 | 4.375 | 4.671 | 5.127 | 5.907 | 7.500 | 12.369 | | |
| 5 | 2.612 | 2.674 | 2.758 | 2.872 | 3.038 | **3.296** | 3.740 | 4.653 | 7.458 | |
| 4 | | | | | | | | 2.166 | 2.625 | 4.051 |

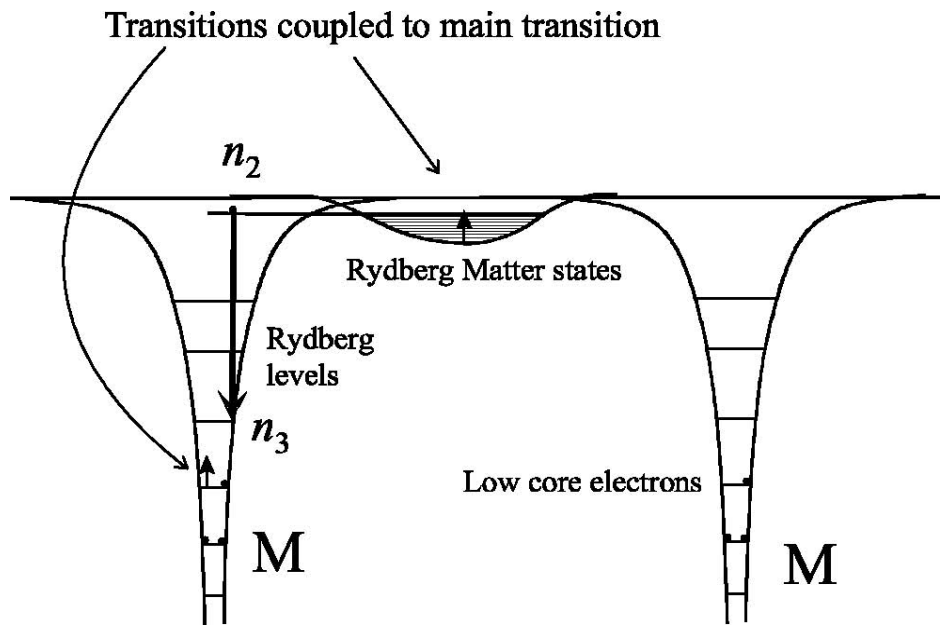

Figure 1. Energy diagram for the main transition $n_2$ to $n_3$ in RM also indicating the secondary transitions coupled to the main transition.

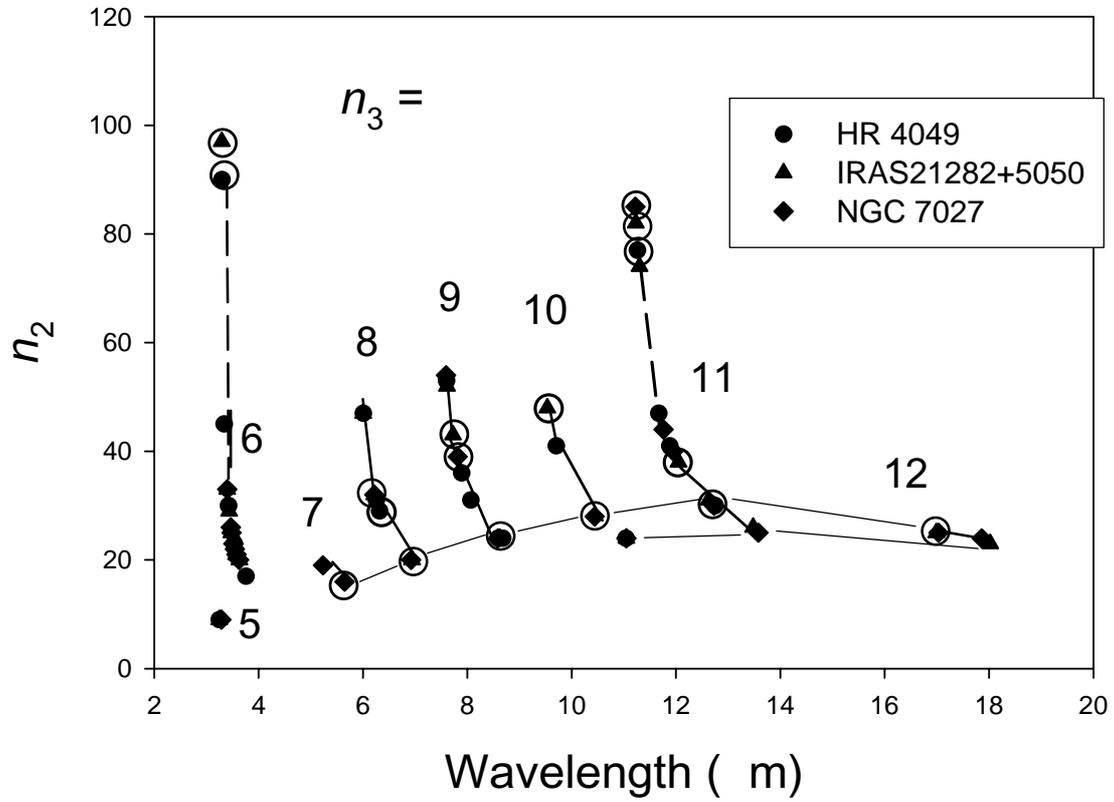

Figure 2. Assignments of the UIR transitions observed in three different stars by Beintema et al. (1996). The values plotted are from Table 1. The very high $n_2$ values for $n_3 = 6$ and 11 are caused by overlap with transitions in Rydberg species, as shown in Table 3. The high intensity bands (bold in Table1) are marked by circles and with a trend indicated.

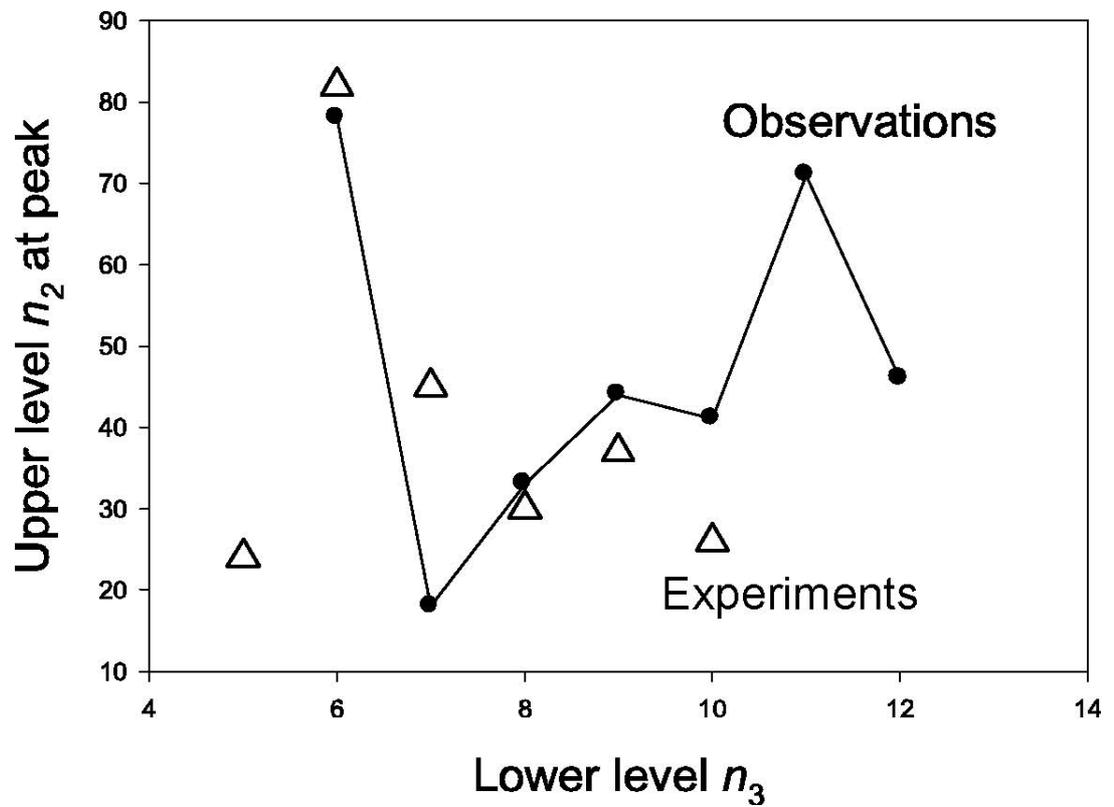

Figure 3. Upper level $n_2$ in Eq. (3) plotted against $n_3$ for experimental results (triangles) (Holmlid 2004b) and observations (circles) (Kahanpää et al. 2003). Note the similarity of the relations from experiments and observation where they overlap. The very high values of $n_2$ at $n_3 = 6$ and 11 are due to enhancement by Rydberg transitions outside RM, as shown in Table 3.